\documentclass[11pt,twoside]{article}

\usepackage{asp2006}
\usepackage{epsf}
\usepackage{psfig}
\usepackage{lscape}

\usepackage{natbib}
\usepackage{graphicx}

\begin{document}
\title{The Properties of Galaxies in 
Low Density Regions from the ALFALFA Survey}   
\author{M.C.~Toribio and J.M.~Solanes for the ALFALFA collaboration}   
\affil{Departament d'Astronomia i Meteorologia and 
Institut de Ci\`encies del Cosmos, Universitat de Barcelona,
Mart\'i i Franqu\`es 1, E-08028 Barcelona, Spain}    

\begin{abstract} 
  Galaxies detected in the 21-cm line of neutral hydrogen (H$\,${\footnotesize I}) from
  the on-going Arecibo Legacy Fast ALFA (ALFALFA) blind extragalactic
  H$\,${\footnotesize I}\ survey have been cross-correlated with Sloan Digital Sky Survey
  (SDSS) Data Release 7 \citep{Aba09} in order to define a
  reference sample of H$\,${\footnotesize I}\ content in regions of low galactic
  density. This observational sample will be used in the future to
  derive new standards of normal atomic gas content that allow a
  statistical investigation of the H$\,${\footnotesize I}\ properties of galaxies in
  differing environments of the local universe. As a previous step, we compare 
  here morphological indicators, like color or light concentration index,
  of ALFALFA detections and non-detections   in low density regions. 
  Our examination is extended also 
  to a small data set of isolated galaxies. This kind of analysis is necessary in order to
  characterize as accurately as possible the type of galaxies that
  ALFALFA is detecting.
\end{abstract}

\section{Introduction}

The first study to establish rigorous standards of normalcy for
the H$\,${\footnotesize I}\ content of galaxies in the absence of external influences was
carried out by \citet{HG84}, who used 288 galaxies with neutral
hydrogen emission contained in the \emph{Catalogue of Isolated
  Galaxies} \citep[CIG,][]{Kar73} to define a control sample. In this
work, it was demonstrated that, for a given Hubble type, the optical
linear diameter of galaxies is the most accurate diagnostic tool for
the H$\,${\footnotesize I}\ mass. This measure of 
the H$\,${\footnotesize I}\ content was reviewed and
updated by \citet*{SGH96} using a sample of 532 
galaxies listed in the \citet{Zwi6168} magnitude-limited
Catalog of Galaxies and Clusters of Galaxies located in the lowest
density environments of the Pisces-Perseus supercluster region. 
The lack of H$\,${\footnotesize I}\ dedicated surveys in those days forced the authors to
deal with multiple galaxy data sets affected by 
incompleteness and selection effects 
that could not always be fully accounted for in the calculation 
of the standard measurements of the H$\,${\footnotesize I}\ content.

This situation is now changing rapidly thanks to the on-going Arecibo
Legacy Fast ALFA (ALFALFA) survey \citep{Gio05}, a blind, flux-limited
extragalactic survey of the sky accesible to the Arecibo's antenna
that is providing an homogeneous census of 21-cm line
sources over a cosmologically significant volume of the local
universe. In this work, we use the H$\,${\footnotesize I}\ dataset from ALFALFA that will
soon constitute the first massive public data release, to define
a standard sample for the comparison of the H$\,${\footnotesize I}\ content of galaxies
free of the potential observational biases mentioned above, and 
combine it with optical data from SDSS in order to inspect
some of the intrinsic properties of ALFALFA sources. 

The definition of a suitable standard of normalcy 
for the H$\,${\footnotesize I}\ content
necessitates a well-chosen control sample formed by galaxies 
with H$\,${\footnotesize I}\
properties as less affected by the environment as possible. This
requirement has led us to define our reference sample from ALFALFA
detections in regions of low galactic density. This procedure, which
is supported by previous works showing that the depletion 
of H$\,${\footnotesize I}\ gas
is a phenomenon essentially constrained to the inner regions of rich
clusters and the densest groups of galaxies 
\citep[e.g.][]{GH85, Sol01, Ver01}, ensures a sufficiently 
large sample that allows a thorough 
statistical treatment of data. 

In the present work, we also
investigate the properties of a much smaller dataset of isolated
galaxies, the objects which are presumably less affected by
environmental processes.

\section{Low Density Region galaxy sample}\label{LDR}

The ALFALFA catalog consists of $\sim$6800 H$\,${\footnotesize I}\ detections over
 $\sim$ 1440 deg$^2$ on the Northern Hemisphere
sky and with H$\,${\footnotesize I} -masses within the range 
$10^7 \leq h^2 M_{HI}/M_{\odot} \leq 10^{10}$.
We have restricted our attention to those ALFALFA galaxies 
which have an optical counterpart in SDSS and heliocentric velocities 
within the range $3000 \leq v_{\mathrm{hel}}\leq 15000$ \ km\ s$^{-1}$, in order
to avoid large distance uncertainties for the
closest objects as well as the gap in 21-cm line detections above
15000 \ km\ s$^{-1}$\ caused by radio frequency interference. For each candidate, 
and after correcting radial distances for peculiar motions following 
prescriptions by \citet{Bla05}, we calculate the 3D distance to the $6th$ nearest
neighbor in the SDSS complete spectroscopic survey (with $r$-band
magnitude $\leq$ 17.77 mag). The corresponding estimate of the local
density $\rho_6$ is corrected for Galactic extinction \citep{SFD98}
and magnitude-limit effects by using the SDSS luminosity function
\citep{Bla01}.

The Low Density Region (LDR) sample includes galaxies with local
density below a certain threshold $\rho_{\mathrm{thr}}$. By applying
the H$\,${\footnotesize I}-deficiency estimator \citep{GH83} updated with the
coefficients inferred in \cite{SGH96}, we find that the H$\,${\footnotesize I}\ content
of galaxies inhabiting regions with local density below
$\rho_{\mathrm{thr}}=0.5$~${\mathrm{galaxies/Mpc}}^3$ is not correlated with
the environment. We have also verified that this threshold allows to
discard practically all galaxies located within $\sim$2-3 virial radii
around rich Abell clusters.

\begin{figure}[t]
\includegraphics[width=\textwidth]{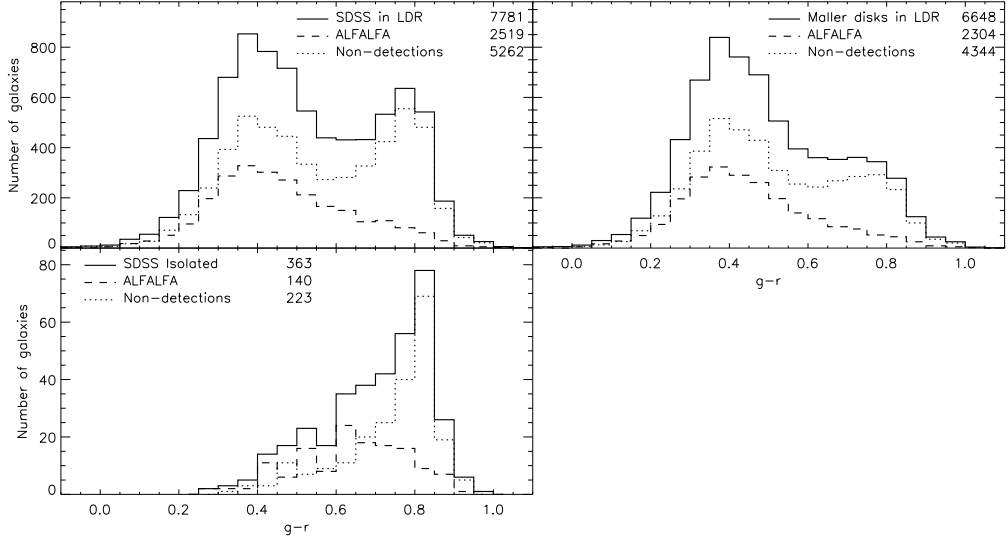}

\caption{\emph{Top left:} SDSS $g-r$ color of a subset of the SDSS
  spectroscopic complete sample in the region overlapping with ALFALFA survey
  (solid). We have selected only galaxies in low density
  regions with velocities within $3000 \leq
  v_{\mathrm{hel}} \leq 15000$ \ km\ s$^{-1}$. The sample is splitted into
  objects whose H$\,${\footnotesize I}\ is detected by ALFALFA (dashed line), and
  non-detections (dotted line). \emph{Top right:} same as above, but only
  for those objects identified as disks according to the
  \citeauthor{Mal09}'s \citeyear{Mal09} criterion.  \emph{Bottom left:}
  same as above, but for objects classified as isolated applying our
  spectrophotometric criterion.}\label{fig1}

\end{figure}

In the top left panel of Figure~\ref{fig1}, we compare the distributions of
$g-r$ colors of SDSS and ALFALFA galaxies located in LDR 
in the volume of the sky (in \emph{z}-space) where both surveys
overlap. This comparison is repeated in the top right panel, but
restricted to the subsets of LDR galaxies that verify the criterion
used by \citet{Mal09} to identify disk galaxies (axis ratio $\leq
0.55$ or S\'ersic index $\leq 3$). As expected, most of ALFALFA
detections ($\sim 92\%$) satisfy it, while about one non-detection out
of every six is not a \citeauthor{Mal09}'s disk.

Inspection of the color distribution confirms that ALFALFA is
detecting mainly blue galaxies. Even among those objects classified as
disks by \citeauthor{Mal09}'s criterion, ALFALFA galaxies are also the
bluest. Given the flux-limited nature of the ALFALFA survey, this
behavior becomes more accentuated with distance, i.e.\ the galaxies
with the reddest colors, which are expected to be usually less rich in
H$\,${\footnotesize I}, are preferently detected at closer distances. 
We also find that ALFALFA detections tend to show large values of the inverse
concentration index, a characteristic of late-type galaxies.

\section{Isolated galaxy sample}\label{isolated}

We have compared the colors of the members of the LDR galaxy sample
defined in Section \ref{LDR} with those belonging to a data set of
isolated galaxies that are as unperturbed as possible by environmental
mechanisms. The latter has been defined applying different isolation
criteria, such as the one based on angular diameter used by
\citet{Kar73} in the CIG and its updated version adopted by
\citet{All05} for the SDSS. We have also defined our own isolation
criterion similar to others found in the literature that combine
photometric and spectroscopic information: a candidate galaxy $i$ is
considered to be isolated if any neighboring galaxy $j$ with apparent
magnitude $m_j \le m_i+1.5$ mag is located further away than 280
$h^{-1}$ kpc ($\sim$ 400 kpc for $h=$0.7).

As already revealed by previous works \citep[e.g.][]{Ver07}, while it
is true that isolated galaxies avoid the densest regions of the
universe, we find that some of them may inhabit regions of moderate
local density in which $\rho_6$ is larger than the density threshold
we have used to define our LDR sample. To avoid this problem, 
membership for our isolated galaxy sample is restricted to LDR objects only.

In sharp contrast with LDR galaxies, dominated by blue objects, the
color distribution of our isolated sample demonstrates that any technique 
used to identify isolated galaxies from the difference in apparent
magnitude between neighbors \emph{always} favours the selection of
brighter and, hence, redder objects (Figure~\ref{fig1}). This is consistent,
for instance, with the results reported in the work by \citet{All05},
in which they obtained an isolated galaxy sample from SDSS with a
concentration index distribution that suggested a morphological
composition fifty-fifty between early and late galaxy types.

On the other hand, we find that, from a statistical point of view, the
 H$\,${\footnotesize I}\ contents of our isolated 
galaxies are essentially identical to
those of our LDR sample galaxies, thus confirming that substantial gas
depletion takes place mainly in high density environments.

\section{Summary and Future Work}

We have introduced the reference data set of ALFALFA
detections that is going to be used in the calculation of new
standards of the H$\,${\footnotesize I}\ content of galaxies.
The present comparison 
between some characteristics of SDSS galaxies in LDR detected 
and undetected by the ALFALFA survey is just a first small step 
towards setting up a suitable 
framework for the investigation of the H$\,${\footnotesize I}\ content of
galaxies in a wide range of environments of the local universe.

\acknowledgements 
This work is supported by the Spanish Direcci\'on
General de Investigaci\'on Cient\'{\i}fica y T\'ecnica, under contract
AYA2007-60366. M.C.T.\ acknowledges support from a fellowship of the
Ministerio de Educaci\'on y Ciencia of Spain.


\begin{thebibliography}{}

\bibitem[{Abazajian} {et~al.}(2009) {Abazajian} {et~al.}]{Aba09}
{Abazajian}, K.~N., {Adelman-McCarthy}, J.~K., {Ag{\"u}eros}, M.~A. 
et al. 2009, \apjs, 182, 543


\bibitem[{Allam} {et~al.}(2005){Allam} {et~al.}]{All05}
{Allam}, S.~S., {Tucker}, D.~L., {Lee}, B.~C., \& {Smith}, J.~A. 2005, \aj,
  129, 2062

\bibitem[{Blanton} {et~al.}(2001){Blanton} {et~al.}]{Bla01}
{Blanton}, M.~R., {Dalcanton}, J., {Eisenstein}, D. et al. 2001, \aj, 121, 2358

\bibitem[{Blanton} {et~al.}(2005){Blanton} {et~al.}]{Bla05}
{Blanton}, M.~R., {Schlegel}, D.~J., {Strauss}, M.~A. {et~al.} 2005, \aj, 129, 2562


\bibitem[{Giovanelli} \& {Haynes}(1983){Giovanelli} \& {Haynes}]{GH83}
{Giovanelli}, R., \& {Haynes}, M.~P. 1983, \aj, 88, 881

\bibitem[{Giovanelli} \& {Haynes}(1985){Giovanelli} \& {Haynes}]{GH85}
---. 1985, \apj, 292, 404

\bibitem[{Giovanelli} {et~al.}(2005){Giovanelli} {et~al.}]{Gio05}
{Giovanelli}, R., {Haynes}, M.~P., {Kent}, B.~R. et al. 2005, \aj, 130, 2598

\bibitem[{Haynes} \& {Giovanelli}(1984){Haynes} \& {Giovanelli}]{HG84}
{Haynes}, M.~P., \& {Giovanelli}, R. 1984, \aj, 89, 758

\bibitem[{Karachentseva}(1973){Karachentseva}]{Kar73}
{Karachentseva}, V.~E. 1973, Soobshcheniya Spetsial'noj Astrofizicheskoj
  Observatorii, 8, 3

\bibitem[{Maller} {et~al.}(2009){Maller}  {et~al.}]{Mal09}
{Maller}, A.~H., {Berlind}, A.~A., {Blanton}, M.~R., \& {Hogg}, D.~W. 2009,
  \apj, 691, 394

\bibitem[{Schlegel} {et~al.}(1998){Schlegel}, {Finkbeiner}, \&
  {Davis}]{SFD98}
{Schlegel}, D.~J., {Finkbeiner}, D.~P., \& {Davis}, M. 1998, \apj, 500, 525

\bibitem[{{Solanes} {et~al.}(1996){Solanes}, {Giovanelli}, \& {Haynes}}]{SGH96}
{Solanes}, J.~M., {Giovanelli}, R., \& {Haynes}, M.~P. 1996, \apj, 461, 609


\bibitem[{Solanes} {et~al.}(2001)]{Sol01}
  {Solanes}, J.~M., {Manrique}, A., {Garc{\'{\i}}a-G{\'o}mez}, C. {et~al.} 2001, \apj, 548, 97

\bibitem[{Verdes-Montenegro} {et~al.}(2001)]{Ver01}
{Verdes-Montenegro}, L., {Yun}, M.~S., {Williams}, B.~A. {et~al.} 2001, \aap, 377,812


\bibitem[{{Verley} {et~al.}(2007){Verley}{et~al.}}]{Ver07}
{Verley}, S., {Leon}, S., {Verdes-Montenegro}, L.  et al. 2007, \aap, 472, 121

\bibitem[Zwicky et al.(1961--1968)]{Zwi6168} Zwicky, F., Herzog, E.,
  Wild, P., Karpowicz, M., \& Kowal, C.\ 1961--1968, Catalog of
  Galaxies and Clusters of Galaxies (Pasadena: California Inst.\ of
  Tech.\ Press)

\end{thebibliography}
\end{document}